\documentclass[aip,apl]{revtex4-1}

\usepackage{amsmath}
\usepackage{mathrsfs}
\usepackage{amssymb}
\usepackage{amsfonts}
\usepackage{graphicx}
\usepackage{float}
\usepackage{epstopdf}
\usepackage[colorlinks=true,linkcolor=blue]{hyperref}%
\expandafter\ifx\csname package@font\endcsname\relax\else
 \expandafter\expandafter
 \expandafter\usepackage
 \expandafter\expandafter
 \expandafter{\csname package@font\endcsname}%
\fi
\begin{document}

\title{Ghost imaging for an occluded object}
\author{Chao Gao}
\affiliation{Department of Physics, Changchun University of Science and Technology,
Changchun 130022, P. R. China}

\author{Xiaoqian Wang}
\email{xqwang21@163.com}
\affiliation{Department of Physics, Changchun University of Science and Technology,
Changchun 130022, P. R. China}

\author{Lidan Gou}
\affiliation{Department of Physics, Changchun University of Science and Technology,
Changchun 130022, P. R. China}

\author{Yuling Feng}
\affiliation{Department of Physics, Changchun University of Science and Technology,
Changchun 130022, P. R. China}

\author{Hongji Cai}
\affiliation{Department of Physics, Changchun University of Science and Technology,
Changchun 130022, P. R. China}

\author{Zhifeng Wang}
\affiliation{Department of Physics, Changchun University of Science and Technology,
Changchun 130022, P. R. China}

\author{Zhihai Yao}
\email{yaozh@cust.edu.cn}
\affiliation{Department of Physics, Changchun University of Science and Technology,
Changchun 130022, P. R. China}

\begin{abstract}
Imaging for an occluded object is usually a difficult problem, in this letter, we introduce an imaging scheme based on computational ghost imaging, which can obtain the image of a target object behind an obstacle. According to our theoretical analysis, once the distance between the object and the obstacle is far enough, one can obtain the image of the object by using ghost imaging technique. The wavelength of the light source also affects the quality of the reconstructed image. In addition, if the bucket detector is placed far away from the obstacle, a tiny point-like detector without collecting lens can be applied to realize the imaging. These theoretical results above have been verified with our numerical simulations. Furthermore, the robustness of this imaging scheme is also investigated.
\end{abstract}

\maketitle
Ghost imaging is a novel imaging technique based on the intensity fluctuation correlations of the light, and it was first proposed with entangled photons\cite{sov,pit}. Later, it was found that ghost imaging could also be realized by using classical thermal source\cite{1stclassical}, and there were many discussions about thermal ghost imaging\cite{thermal1,thermal2,thermal3,Gatti2006,gatti2006a,shih2006,Ferri2008,thermal4,thermalA,visi,higi,wulingan1,wulingan2,Gao2017}. \par
In 2008, J.H.Shapiro proposed computational ghost imaging\cite{cgi}, and it was verified by experiment in 2009\cite{cgie}. Different from conventional ghost imaging scheme, computational ghost imaging technique applies a programmable light source, and the experimental setup can be simplified. Ghost imaging displays great potentials in some special situations, such as high lateral resolution imaging\cite{superresolution}, resistance of atmosphere turbulence\cite{turbulence,turbulencei} and so on. \par

In addition to the above features, our recent work shows that ghost imaging may have even more advantages than conventional imaging techniques. Imaging for an occluded object is a difficult problem, in this letter, we proved that, under appropriate condition, one can obtain the image of an occluded object by applying ghost imaging technique, even if the object is blocked by an unknown obstacle. \par

\begin{figure}[h]
\fbox{\includegraphics[width=\linewidth]{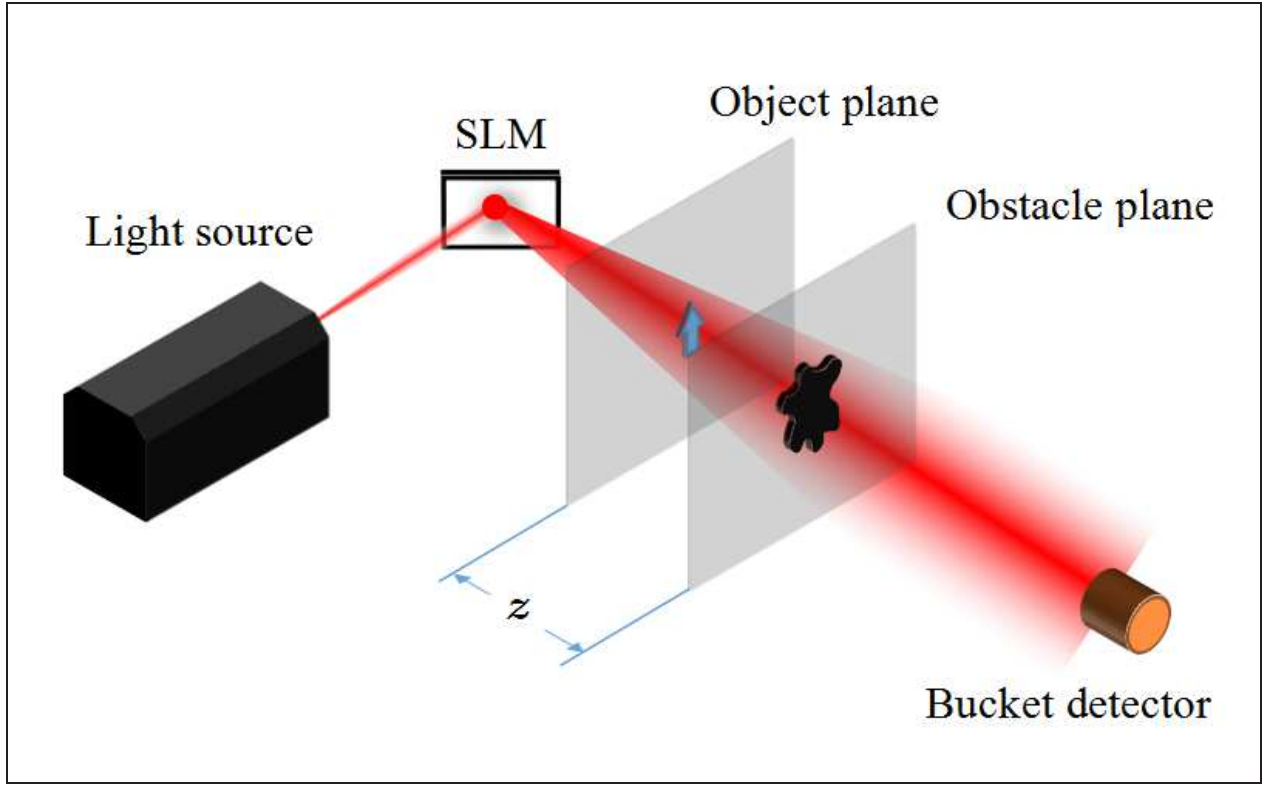}}
\centering
\caption{The schematic diagram of the computational ghost imaging for an occluded object. The distance between the target object plane and the obstacle plane is $z$.}
\label{ns}
\end{figure}

The schematic diagram of the computational ghost imaging for an occluded object is shown in Fig.~\ref{ns}. If we view from the bucket detector, the target object is blocked by the obstacle. But when we use computational ghost imaging technique, we can obtain the object's image. \par
For simplification, we consider the 1-dimension case. Let $\vec C$ and $\vec D$ represent the transmission functions of the target object and the obstacle, where $\vec C=[c_1,c_2,\cdots,c_N]^T$ and $\vec D=[d_1,d_2,\cdots,d_N]^T$. The light emitted by the programmable light source illuminates the object, and its intensity distribution on the object plane can also be represented by an $1\times N$ vector:
\begin{equation}
\vec S(t)=
\left[
\begin{array}{cccccc}
s_1(t)&s_2(t)&...&s_n(t)&...&s_N(t)\\
\end{array}
\right] ^T.
\end{equation}
The modulated light illuminates and passes through the target object, reached the obstacle plane after $z$ distance of propagating. First, for simplification, we assume that we can collect all the transmitted light by using a bucket detector, and the bucket signal can be written as:
\begin{equation}
B(t)=\sum\limits_{m=1}^{N}d_m\sum\limits_{n=1}^{N}A_{mn}c_ns_n(t).
\end{equation}
which $\{c_n\}$ and $\{d_m\}$ represent the elements of $\vec C$ and $\vec D$. $\{A_{mn}\}$ represent the elements of the propagating matrix $\hat A$. The second-order correlation function\cite{quantumoptics} of this system is:
\begin{equation}
\vec G^{(2)}(n')=\langle \vec s_{n'}(t) B(t)\rangle_t.
\label{g2o}
\end{equation}
Here, $\langle ...\rangle_t$ represents the ensemble average. In this system, we assume that $\{s_n(t)\}$ are independent and identically distributed. If we take enough measurements, we have\cite{quantumoptics}:
\begin{equation}
\langle s_{n'}(t)s_n(t) \rangle_t=\delta(n',n)l+\langle s \rangle^2.
\end{equation}
Where $\langle s\rangle$ is the average intensity of the light source, $l$ is the variance of the intensity. Therefore, the non-normalized second-order correlation function of the target object can be expressed by:
\begin{equation}
\vec G^{(2)}(n')=lc_{n'}\sum\limits_{m=1}^{N}d_mA_{mn'}+O.
\label{result1}
\end{equation}
\par
Where $O=\langle s\rangle^2\sum\limits_{m=1}^{N}d_m\sum\limits_{n=1}^{N}A_{mn}c_n$ is a background term which is unrelated to $n'$. In another word, it does not contains the spatial information of the object. So, we focus on term $lc_{n'}\sum\limits_{p=1}^{N}d_mA_{mn'}$. Obviously, $\vec G^{(2)}$ is in proportion to $c_{n'}\sum\limits_{m=1}^{N}d_mA_{mn'}$. In order to find out the relationship between $\vec G^{(2)}$ and the transmission function of the target object, we need to investigate the form of the propagating matrix $\hat A$. \par
\begin{figure}[h]
\fbox{\includegraphics[width=\linewidth]{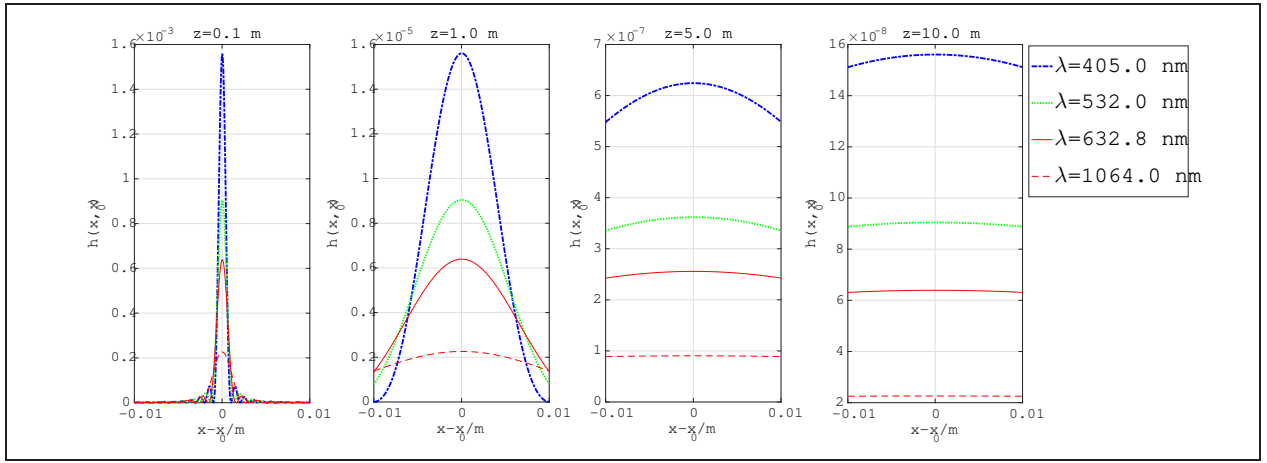}}
\centering
\caption{The intensity point spread function of the light sources with the wavelengthes of $405nm$, $532nm$, $632.8nm$ and $1064nm$ at different distances. Where $x-x_0=(m-n)\Delta x$. }
\label{psf}
\end{figure}

Now, we investigate the propagation progress, and we consider the dispersed case. Let $u_0(n,t')$ be the instantaneous field distribution of the source on the target object plane at time $t'$. For simplification, the dimension and the pixel size at the target object and obstacle planes are equal. For the pixel with transverse size of $\Delta x$, after $z$ distance of traveling, the field distribution on the obstacle plane can be written as\cite{fddt}:
\begin{equation}
u(m,t')=\frac{e^{ikz}(\Delta x)^2}{i\lambda z}\sum\limits_{n=1}^{N}u_0(n,t')e^{\frac{ik}{2z}(m-n)^2(\Delta x)^2}.
\end{equation}
Where N is the number of the pixels, and $\frac{e^{ikz}(\Delta x)^2}{i\lambda z}\sum\limits_{m=1}^{N}e^{\frac{ik}{2z}(m-n)^2(\Delta x)^2}$ is usually denoted by $h_{z,\lambda}(m-n)$, which is called the point spread function (PSF). We can obtain the intensity distribution on the obstacle plane:
\begin{align}
&I_{obs}(m,t)=\int_{t}^{t+\Delta t} u^*(m,t')u(m,t')dt' \notag \\
&=\int_{t}^{t+\Delta t}[h^*(m-1)u_0^*(1,t')+h^*(m-2)u_0^*(2,t')+... \notag \\
&+h^*(m-n)u_0^*(n,t')+...h^*(m-N)u_0^*(N,t')]\times \notag \\
&[h(m-1)u_0(1,t')+h(m-2)u_0(2,t')+... \notag \\
&+h(m-n)u_0(n,t')+...+h(m-N)u_0(N,t')]dt'.
\label{I2}
\end{align}
Where $\Delta t$ is a short period of time. Here, we assume that the light source is incoherent. In this case, we have:
\begin{equation}
\int_{t}^{t+\Delta t} u^*(x_1,t')u(x_2,t') dt'=\delta(x_1,x_2)|u(x_1,t)|^2.
\end{equation}
And we can simplify Eq.~(\ref{I2}) into:
\begin{equation}
I_{obs}(m,t)=\sum\limits_{n'=1}^{N}\sum\limits_{n=1}^{N}h^*(m-n')h(m-n)I_{obj}(n,t).
\label{obs}
\end{equation}
Where $I_{obj}(n,t)$ is the intensity distribution through the target object plane. Eq.~(\ref{obs}) can also be written in the matrix form:
\begin{equation}
I_{obs}=
\left[
\begin{array}{cccccc}
A_{11} &A_{12} &\cdots &A_{1n} &\cdots &A_{1N} \\
A_{21} &A_{22} &\cdots &A_{2n} &\cdots &A_{2N} \\
\vdots &\vdots &\ddots &\vdots &       &\vdots \\
A_{m1} &A_{m2} &\cdots &A_{mn} &\cdots &A_{mN} \\
\vdots &\vdots &       &\vdots &\ddots &\vdots \\
A_{N1} &A_{N2} &\cdots &A_{Nn} &\cdots &A_{NN}
\end{array}
\right]I_{obj}.
\end{equation}
Where
\begin{equation}
A_{mn}=|h(m-n)|^2=\frac{(\Delta x)^4}{\lambda^2 z^2}\sum\limits_{n'=1}^{N}e^{i\frac{\pi (\Delta x)^2 }{\lambda z}[(m-n')^2-(m-n)^2]}
\label{matrixa}
\end{equation}
is called the intensity point spread function. Fig.~\ref{psf} gives the intensity point spread function curve of light sources with several typical wavelengthes at different distances of propagating. \par
Fig.~\ref{psf} shows that the intensity point spread function is influenced by the distance of propagating $z$ and the wavelength of the light source $\lambda$. When $\lambda$ and/or $z$ is big enough, the intensity point spread function approach to a constant which is unrelated to the spatial coordinates. \par
The physics behind this progress is: the illuminating light carries the information of the object, propagates a distance of $z$, and reach the obstacle plane. Due to the propagation of the light, the information of the object spread around on the obstacle plane. Every single point on the object plane produces an Airy pattern on the obstacle plane, and the Airy patterns overlap with each other. As a result, every single pixel on the obstacle plane contains the information from multiple points on the object plane. As distance $z$ or wavelength $\lambda$ increases, the area of every Airy pattern increases. While the magnitudes of $z$ and/or $\lambda$ are great enough, we can assume that every pixel on the obstacle plane contains the information from all of the points on the object plane. So that the effective information of the object can always reach the bucket detector via the outside of the obstacle's border. \par
The diffraction on the obstacle plane is actually a similar progress: after a distance of traveling, the transmitted light reaches the bucket detector plane. Noticed that, like the situation we discussed above, if this distance is far enough, every pixel on the bucket detector plane contains the information from all of the points on the obstacle plane. Therefore, in this case, it is not necessary to collect all of the transmitted light. Instead, in ideal condition, even a tiny point-like detector can finish the task. \par
With the assumption of a long distance between the bucket detector and the obstacle, the intensity on the bucket detector plane approaches to be evenly distributed. The bucket signal can be written as:
\begin{equation}
B'(t)=\alpha \sum\limits_{m=1}^{N}d_m\sum\limits_{n=1}^{N}A_{mn}c_ns_n(t)
\end{equation}
Where $\alpha \in (0,1)$ is a constant which depends on the size of the bucket detector.
Based on the discussions above, we can now explain why ghost imaging technique can realize the imaging of an occluded object. From Eq.~(\ref{result1}) we know that $\vec G^{(2)}$ is in proportion to $c_{n'}\sum\limits_{m=1}^{N}d_mA_{mn'}$. Both the information of the target object and the obstacle are contained in $\vec G^{(2)}$. While the magnitude of $\lambda z$ is big enough, the elements in the propagating matrix $\hat A$ approach to a constant which is unrelated to the spatial coordinates. The second-order correlation function of the target object is:
\begin{equation}
\vec G^{(2)}(n')\approx l\alpha\bar{d}\bar{A}c_{n'}+\alpha O.
\end{equation}
 Where $\bar{d}=\sum\limits_{n=1}^{N}d_n$, $\bar{A}=\frac{1}{N}\sum\limits_{m=1}^{N}\sum\limits_{n=1}^{N}A_{mn}$. In this case, the spatial information of the obstacle is eliminated. $G^{(2)}$ is now in proportion to the target object's transmission function, the image of the object can be obtained correctly. Noticed that, one can obtain the image of the target object in this case, even if the shape of the obstacle is unknown. The reason is, different from conventional imaging technique, ghost imaging is a kind of computational imaging scheme which is based on the intensity fluctuation correlations, the imaging quality is only sensitive to the fluctuation of the total (or average) intensity of the transmitted light. When the distance between the object and the obstacle is far enough, the obstacle does very limited effects on the fluctuation of the bucket signal. The result is: in this case, even under the affect of an obstacle, ghost imaging scheme will not fail, we can still obtain the image of the target object. \par

However, when $\lambda z$ decreases, the curve of the intensity point spread function approaches to $\delta$ function. Thus, the non-opposite angle elements of propagating matrix $\hat A$ approach to zero. In this case, the second-order correlation function of the target object can be written as:
\begin{equation}
\vec G^{(2)}(n')\approx l\alpha A_{n'n'}c_{n'}d_{n'}+\alpha O.
\end{equation}
\par
Obviously, $\vec G^{(2)}$ is in proportion to $c_{n'}d_{n'}$, the product of the transmission function of the target object and the obstacle. We will obtain the mixture image of the target object and the obstacle, we cannot revive the image of the target object correctly. \par

Thus, to realize the imaging for an occluded object, the distance between the target object and the obstacle should be far enough. Besides, in order to obtain a image with higher quality, we can increase the wavelength of the illuminating light. Furthermore, if we place the bucket detector far away from the obstacle, it is possible to use a tiny detector to realize the imaging.\par

\begin{figure}[h]
\fbox{\includegraphics[width=6 cm]{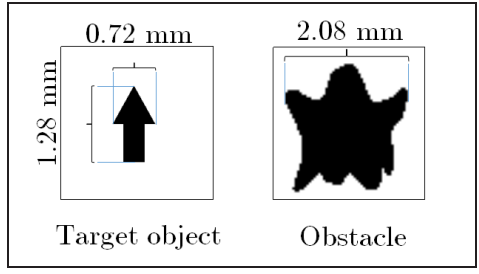}}
\centering
\caption{The sizes of the target object and the obstacle.}
\label{objects}
\end{figure}

To verify our theoretical results, the numerical simulations are carried out, and, the robustness of this imaging system is also judged. The schematic diagram of our numerical simulation is shown in Fig.~\ref{ns}. We take $1,200,000$ measurements for every simulation, and the field distribution of light source is modulated into gaussian randomly distributed. The distance between SLM and the target object is taken to be $0.50$m. The size of the bucket detector is $0.08\times 0.08$mm, and it is placed $10.00$m far from the obstacle, in the center of the bucket detector plane (on the optic axis). As Fig.~\ref{objects} shows, the target object is an opaque arrow, and the obstacle is a ``ghost''-shaped opaque plate, both of them are placed in the center in their planes. The size of the target object is $1.20\times 0.72$ mm, and the size of the obstacle is about $2.08 \times 2.08$ mm. Both the target object and the obstacle plane are pixelated into two $64\times 64$ pixels images, with pixel width $\Delta x=0.04$ mm.
\par We investigate the influence of the distance between the target object and the obstacle and the wavelength of the light source, respectively.
\subsection{The influence of the distance between the target object and the obstacle}
\begin{figure}[h]
\fbox{\includegraphics[width=6 cm]{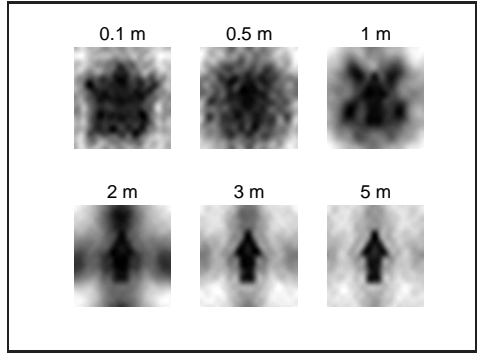}}
\centering
\caption{The reconstructed images of the target object for different distances between the target object and the obstacle. The wavelength of the light source is taken to be $632.8$ nm, and the distances between the target object and the obstacle are $0.1$ m, $0.5$ m, $1.0$ m, $2.0$ m, $3.0$ m and $5.0$ m, respectively.}
\label{distances}
\end{figure}

In this part, we use a $632.8$ nm laser as the light source.
In order to study the influence on the imaging quality of the target object, we change the distance between the target object and the obstacle, and reconstruct the image of the target object by using computational ghost imaging technique, respectively. The results of our numerical simulation are shown in Fig.~\ref{distances}.
It is clear that, while the distance between the target object and the obstacle is far enough, it is possible to realize the imaging for an occluded object by applying computational ghost imaging technique.

\subsection{The influence of the wavelength of the light source}
\begin{figure}[h]
\fbox{\includegraphics[width=6 cm]{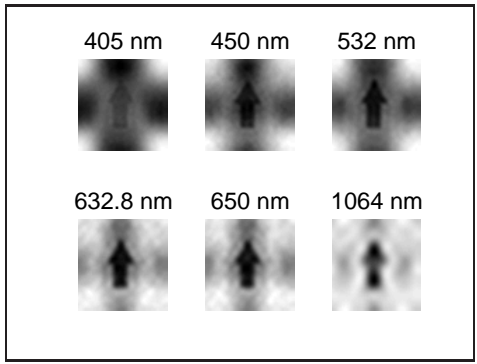}}
\centering
\caption{The reconstructed images of the target object by using light sources with different wavelengthes. The distance between the target object and the obstacle is $3.0$ m, and the wavelengthes are $405.0$ nm, $450.0$ nm, $532.0$ nm, $632.8$ nm, $650.0$ nm and $1064.0$ nm, respectively.}
\label{lambda}
\end{figure}

In this part, the distance between the target object and the obstacle is taken to be $3.0$ m. In order to find out the influences of the wavelength on the imaging quality, we use light sources with different wavelengthes to implement the computational ghost imaging for the target object. The results of our numerical simulations are shown in Fig.~\ref{lambda}.  \par
Obviously, we can get a clearer view of the target object by applying a light source with longer wavelength. However, the spatial resolution of the reconstructed image is decreased. \par
\subsection{The robustness of this imaging system}
\begin{figure}[h]
\fbox{\includegraphics[width=\linewidth]{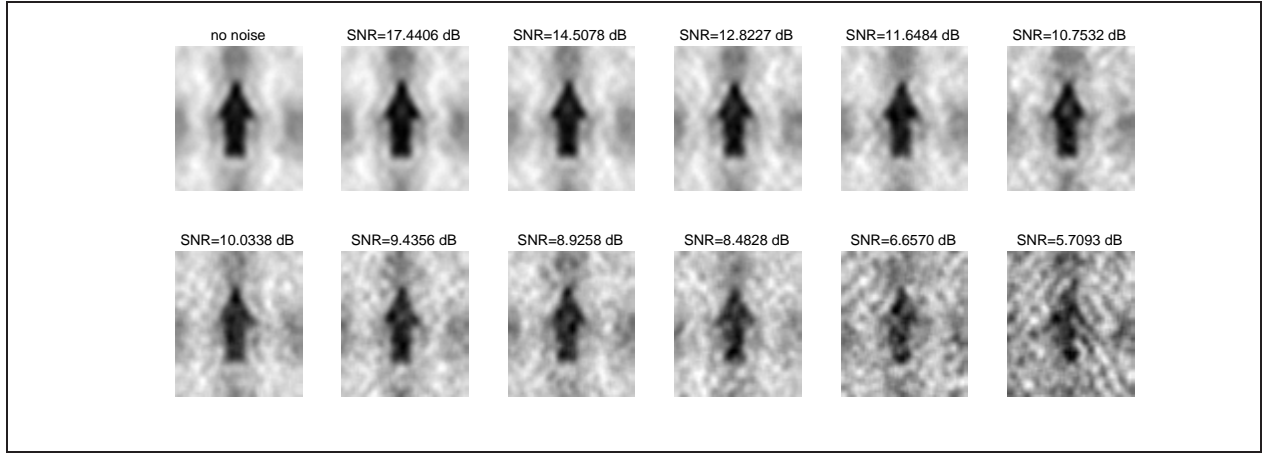}}
\centering
\caption{The reconstructed images of the target object under different detection SNR. The distance between the target object and the obstacle is $3.0$ m, and the wavelength of the light source is $632.8$ nm.}
\label{snr}
\end{figure}

Many computational imaging schemes fail with the affect of noise, thus it is necessary to judge the performance of our imaging scheme under the influence of background noise. We use signal to noise ratio (SNR) to describe the effect of the background noise on the bucket signal, which is defined as:
\begin{equation}
\text{SNR}=10\log_{10}\frac{\bar B}{\bar N_b},
\end{equation}
where $\bar B$ is the average intensity of the bucket signal, $\bar N_b$ is the average intensity of the background noise, the noise is gaussian noise. The reconstructed images of the target object under different SNR are shown in Fig.~\ref{snr}.\par
The results show that, when SNR of the bucket signal is $6.6570 \text{dB}$, the image of the object can still be recognized, in this case, the average intensity of the noise reached about 22\%. The imaging scheme fails when SNR is lower than $5.7093$ dB (namely with about 27\% noise). Thus, this imaging scheme can partly resist the effect of noise.\par
In conclusion, we have proved that ghost imaging can realize the imaging for an occluded object. According to the above discussions, we find that, this unique feature is based on the fact that ghost imaging technique is based on the intensity fluctuation correlations. Due to the diffraction of the light, the bucket detector (with limited size) can always capture the useful fluctuation information which is caused by different illuminating patterns passes through the object. If distance between the target object and the obstacle is far enough, the image of the target object can be reconstructed accurately. While the target object is close to the obstacle, we will obtain the mixture image of the target object and the obstacle, ghost imaging failed in this case. Besides, a better image of the target object can be obtained by using a light source with longer wavelength, but the resolution of the reconstructed image is decreased. In addition, it is possible to realize the imaging by using a tiny point-like bucket detector if the detector is placed far away from the obstacle. The numerical simulations have been carried out, and the results agree with our theoretical analysis. \par
\section*{Acknowledgement}
This work is supported by:
National Natural Science Foundation of China (11305020); The Science and Technology Research Projects of the Education Department of Jilin Province, China (2016-354);
Natural Science Foundation of Jilin Province(20180520165JH). \par


\begin{thebibliography}{24}%
\makeatletter
\providecommand \@ifxundefined [1]{%
 \@ifx{#1\undefined}
}%
\providecommand \@ifnum [1]{%
 \ifnum #1\expandafter \@firstoftwo
 \else \expandafter \@secondoftwo
 \fi
}%
\providecommand \@ifx [1]{%
 \ifx #1\expandafter \@firstoftwo
 \else \expandafter \@secondoftwo
 \fi
}%
\providecommand \natexlab [1]{#1}%
\providecommand \enquote  [1]{``#1''}%
\providecommand \bibnamefont  [1]{#1}%
\providecommand \bibfnamefont [1]{#1}%
\providecommand \citenamefont [1]{#1}%
\providecommand \href@noop [0]{\@secondoftwo}%
\providecommand \href [0]{\begingroup \@sanitize@url \@href}%
\providecommand \@href[1]{\@@startlink{#1}\@@href}%
\providecommand \@@href[1]{\endgroup#1\@@endlink}%
\providecommand \@sanitize@url [0]{\catcode `\\12\catcode `\$12\catcode
  `\&12\catcode `\#12\catcode `\^12\catcode `\_12\catcode `\%12\relax}%
\providecommand \@@startlink[1]{}%
\providecommand \@@endlink[0]{}%
\providecommand \url  [0]{\begingroup\@sanitize@url \@url }%
\providecommand \@url [1]{\endgroup\@href {#1}{\urlprefix }}%
\providecommand \urlprefix  [0]{URL }%
\providecommand \Eprint [0]{\href }%
\providecommand \doibase [0]{http://dx.doi.org/}%
\providecommand \selectlanguage [0]{\@gobble}%
\providecommand \bibinfo  [0]{\@secondoftwo}%
\providecommand \bibfield  [0]{\@secondoftwo}%
\providecommand \translation [1]{[#1]}%
\providecommand \BibitemOpen [0]{}%
\providecommand \bibitemStop [0]{}%
\providecommand \bibitemNoStop [0]{.\EOS\space}%
\providecommand \EOS [0]{\spacefactor3000\relax}%
\providecommand \BibitemShut  [1]{\csname bibitem#1\endcsname}%
\let\auto@bib@innerbib\@empty
\bibitem [{\citenamefont {Klyshko}(1988)}]{sov}%
  \BibitemOpen
  \bibfield  {author} {\bibinfo {author} {\bibfnamefont {D.}~\bibnamefont
  {Klyshko}},\ }\href@noop {} {\bibfield  {journal} {\bibinfo  {journal}
  {Physics Letters A}\ }\textbf {\bibinfo {volume} {132}},\ \bibinfo {pages}
  {299} (\bibinfo {year} {1988})}\BibitemShut {NoStop}%
\bibitem [{\citenamefont {Pittman}\ \emph {et~al.}(1995)\citenamefont
  {Pittman}, \citenamefont {Shih}, \citenamefont {Strekalov},\ and\
  \citenamefont {Sergienko}}]{pit}%
  \BibitemOpen
  \bibfield  {author} {\bibinfo {author} {\bibfnamefont {T.}~\bibnamefont
  {Pittman}}, \bibinfo {author} {\bibfnamefont {Y.}~\bibnamefont {Shih}},
  \bibinfo {author} {\bibfnamefont {D.}~\bibnamefont {Strekalov}}, \ and\
  \bibinfo {author} {\bibfnamefont {A.}~\bibnamefont {Sergienko}},\ }\href@noop
  {} {\bibfield  {journal} {\bibinfo  {journal} {Physical Review A}\ }\textbf
  {\bibinfo {volume} {52}},\ \bibinfo {pages} {R3429} (\bibinfo {year}
  {1995})}\BibitemShut {NoStop}%
\bibitem [{\citenamefont {Bennink}, \citenamefont {Bentley},\ and\
  \citenamefont {Boyd}(2002)}]{1stclassical}%
  \BibitemOpen
  \bibfield  {author} {\bibinfo {author} {\bibfnamefont {R.~S.}\ \bibnamefont
  {Bennink}}, \bibinfo {author} {\bibfnamefont {S.~J.}\ \bibnamefont
  {Bentley}}, \ and\ \bibinfo {author} {\bibfnamefont {R.~W.}\ \bibnamefont
  {Boyd}},\ }\href@noop {} {\bibfield  {journal} {\bibinfo  {journal} {Physical
  review letters}\ }\textbf {\bibinfo {volume} {89}},\ \bibinfo {pages}
  {113601} (\bibinfo {year} {2002})}\BibitemShut {NoStop}%
\bibitem [{\citenamefont {Gatti}\ \emph
  {et~al.}(2004{\natexlab{a}})\citenamefont {Gatti}, \citenamefont {Brambilla},
  \citenamefont {Bache},\ and\ \citenamefont {Lugiato}}]{thermal1}%
  \BibitemOpen
  \bibfield  {author} {\bibinfo {author} {\bibfnamefont {A.}~\bibnamefont
  {Gatti}}, \bibinfo {author} {\bibfnamefont {E.}~\bibnamefont {Brambilla}},
  \bibinfo {author} {\bibfnamefont {M.}~\bibnamefont {Bache}}, \ and\ \bibinfo
  {author} {\bibfnamefont {L.~A.}\ \bibnamefont {Lugiato}},\ }\href@noop {}
  {\bibfield  {journal} {\bibinfo  {journal} {Physical review letters}\
  }\textbf {\bibinfo {volume} {93}},\ \bibinfo {pages} {093602} (\bibinfo
  {year} {2004}{\natexlab{a}})}\BibitemShut {NoStop}%
\bibitem [{\citenamefont {Gatti}\ \emph
  {et~al.}(2004{\natexlab{b}})\citenamefont {Gatti}, \citenamefont {Brambilla},
  \citenamefont {Bache},\ and\ \citenamefont {Lugiato}}]{thermal2}%
  \BibitemOpen
  \bibfield  {author} {\bibinfo {author} {\bibfnamefont {A.}~\bibnamefont
  {Gatti}}, \bibinfo {author} {\bibfnamefont {E.}~\bibnamefont {Brambilla}},
  \bibinfo {author} {\bibfnamefont {M.}~\bibnamefont {Bache}}, \ and\ \bibinfo
  {author} {\bibfnamefont {L.~A.}\ \bibnamefont {Lugiato}},\ }\href@noop {}
  {\bibfield  {journal} {\bibinfo  {journal} {Physical Review A}\ }\textbf
  {\bibinfo {volume} {70}},\ \bibinfo {pages} {013802} (\bibinfo {year}
  {2004}{\natexlab{b}})}\BibitemShut {NoStop}%
\bibitem [{\citenamefont {Valencia}\ \emph {et~al.}(2005)\citenamefont
  {Valencia}, \citenamefont {Scarcelli}, \citenamefont {D¡¯Angelo},\ and\
  \citenamefont {Shih}}]{thermal3}%
  \BibitemOpen
  \bibfield  {author} {\bibinfo {author} {\bibfnamefont {A.}~\bibnamefont
  {Valencia}}, \bibinfo {author} {\bibfnamefont {G.}~\bibnamefont {Scarcelli}},
  \bibinfo {author} {\bibfnamefont {M.}~\bibnamefont {D¡¯Angelo}}, \ and\
  \bibinfo {author} {\bibfnamefont {Y.}~\bibnamefont {Shih}},\ }\href@noop {}
  {\bibfield  {journal} {\bibinfo  {journal} {Physical review letters}\
  }\textbf {\bibinfo {volume} {94}},\ \bibinfo {pages} {063601} (\bibinfo
  {year} {2005})}\BibitemShut {NoStop}%
\bibitem [{\citenamefont {Gatti}\ \emph {et~al.}(2006)\citenamefont {Gatti},
  \citenamefont {Bache}, \citenamefont {Magatti}, \citenamefont {Brambilla},
  \citenamefont {Ferri},\ and\ \citenamefont {Lugiato}}]{Gatti2006}%
  \BibitemOpen
  \bibfield  {author} {\bibinfo {author} {\bibfnamefont {A.}~\bibnamefont
  {Gatti}}, \bibinfo {author} {\bibfnamefont {M.}~\bibnamefont {Bache}},
  \bibinfo {author} {\bibfnamefont {D.}~\bibnamefont {Magatti}}, \bibinfo
  {author} {\bibfnamefont {E.}~\bibnamefont {Brambilla}}, \bibinfo {author}
  {\bibfnamefont {F.}~\bibnamefont {Ferri}}, \ and\ \bibinfo {author}
  {\bibfnamefont {L.}~\bibnamefont {Lugiato}},\ }\href@noop {} {\bibfield
  {journal} {\bibinfo  {journal} {Journal of Modern Optics}\ }\textbf {\bibinfo
  {volume} {53}},\ \bibinfo {pages} {739} (\bibinfo {year} {2006})}\BibitemShut
  {NoStop}%
\bibitem [{\citenamefont {Bache}\ \emph {et~al.}(2006)\citenamefont {Bache},
  \citenamefont {Magatti}, \citenamefont {Ferri}, \citenamefont {Gatti},
  \citenamefont {Brambilla},\ and\ \citenamefont {Lugiato}}]{gatti2006a}%
  \BibitemOpen
  \bibfield  {author} {\bibinfo {author} {\bibfnamefont {M.}~\bibnamefont
  {Bache}}, \bibinfo {author} {\bibfnamefont {D.}~\bibnamefont {Magatti}},
  \bibinfo {author} {\bibfnamefont {F.}~\bibnamefont {Ferri}}, \bibinfo
  {author} {\bibfnamefont {A.}~\bibnamefont {Gatti}}, \bibinfo {author}
  {\bibfnamefont {E.}~\bibnamefont {Brambilla}}, \ and\ \bibinfo {author}
  {\bibfnamefont {L.~A.}\ \bibnamefont {Lugiato}},\ }\href@noop {} {\bibfield
  {journal} {\bibinfo  {journal} {Physical Review A}\ }\textbf {\bibinfo
  {volume} {73}},\ \bibinfo {pages} {053802} (\bibinfo {year}
  {2006})}\BibitemShut {NoStop}%
\bibitem [{\citenamefont {Scarcelli}, \citenamefont {Berardi},\ and\
  \citenamefont {Shih}(2006)}]{shih2006}%
  \BibitemOpen
  \bibfield  {author} {\bibinfo {author} {\bibfnamefont {G.}~\bibnamefont
  {Scarcelli}}, \bibinfo {author} {\bibfnamefont {V.}~\bibnamefont {Berardi}},
  \ and\ \bibinfo {author} {\bibfnamefont {Y.}~\bibnamefont {Shih}},\
  }\href@noop {} {\bibfield  {journal} {\bibinfo  {journal} {Physical review
  letters}\ }\textbf {\bibinfo {volume} {96}},\ \bibinfo {pages} {063602}
  (\bibinfo {year} {2006})}\BibitemShut {NoStop}%
\bibitem [{\citenamefont {Ferri}\ \emph {et~al.}(2008)\citenamefont {Ferri},
  \citenamefont {Magatti}, \citenamefont {Sala},\ and\ \citenamefont
  {Gatti}}]{Ferri2008}%
  \BibitemOpen
  \bibfield  {author} {\bibinfo {author} {\bibfnamefont {F.}~\bibnamefont
  {Ferri}}, \bibinfo {author} {\bibfnamefont {D.}~\bibnamefont {Magatti}},
  \bibinfo {author} {\bibfnamefont {V.}~\bibnamefont {Sala}}, \ and\ \bibinfo
  {author} {\bibfnamefont {A.}~\bibnamefont {Gatti}},\ }\href@noop {}
  {\bibfield  {journal} {\bibinfo  {journal} {Applied Physics Letters}\
  }\textbf {\bibinfo {volume} {92}},\ \bibinfo {pages} {261109} (\bibinfo
  {year} {2008})}\BibitemShut {NoStop}%
\bibitem [{\citenamefont {Chan}, \citenamefont {O¡¯Sullivan},\ and\
  \citenamefont {Boyd}(2009)}]{thermal4}%
  \BibitemOpen
  \bibfield  {author} {\bibinfo {author} {\bibfnamefont {K.~W.~C.}\
  \bibnamefont {Chan}}, \bibinfo {author} {\bibfnamefont {M.~N.}\ \bibnamefont
  {O¡¯Sullivan}}, \ and\ \bibinfo {author} {\bibfnamefont {R.~W.}\
  \bibnamefont {Boyd}},\ }\href@noop {} {\bibfield  {journal} {\bibinfo
  {journal} {Physical Review A}\ }\textbf {\bibinfo {volume} {79}},\ \bibinfo
  {pages} {033808} (\bibinfo {year} {2009})}\BibitemShut {NoStop}%
\bibitem [{\citenamefont {Chen}\ \emph {et~al.}(2009)\citenamefont {Chen},
  \citenamefont {Liu}, \citenamefont {Luo},\ and\ \citenamefont
  {Wu}}]{thermalA}%
  \BibitemOpen
  \bibfield  {author} {\bibinfo {author} {\bibfnamefont {X.-H.}\ \bibnamefont
  {Chen}}, \bibinfo {author} {\bibfnamefont {Q.}~\bibnamefont {Liu}}, \bibinfo
  {author} {\bibfnamefont {K.-H.}\ \bibnamefont {Luo}}, \ and\ \bibinfo
  {author} {\bibfnamefont {L.-A.}\ \bibnamefont {Wu}},\ }\href@noop {}
  {\bibfield  {journal} {\bibinfo  {journal} {Optics letters}\ }\textbf
  {\bibinfo {volume} {34}},\ \bibinfo {pages} {695} (\bibinfo {year}
  {2009})}\BibitemShut {NoStop}%
\bibitem [{\citenamefont {Chan}, \citenamefont {O¡¯Sullivan},\ and\
  \citenamefont {Boyd}(2010)}]{visi}%
  \BibitemOpen
  \bibfield  {author} {\bibinfo {author} {\bibfnamefont {K.~W.~C.}\
  \bibnamefont {Chan}}, \bibinfo {author} {\bibfnamefont {M.~N.}\ \bibnamefont
  {O¡¯Sullivan}}, \ and\ \bibinfo {author} {\bibfnamefont {R.~W.}\
  \bibnamefont {Boyd}},\ }\href@noop {} {\bibfield  {journal} {\bibinfo
  {journal} {Optics express}\ }\textbf {\bibinfo {volume} {18}},\ \bibinfo
  {pages} {5562} (\bibinfo {year} {2010})}\BibitemShut {NoStop}%
\bibitem [{\citenamefont {Chan}, \citenamefont {O'Sullivan},\ and\
  \citenamefont {Boyd}(2009)}]{higi}%
  \BibitemOpen
  \bibfield  {author} {\bibinfo {author} {\bibfnamefont {K.~W.~C.}\
  \bibnamefont {Chan}}, \bibinfo {author} {\bibfnamefont {M.~N.}\ \bibnamefont
  {O'Sullivan}}, \ and\ \bibinfo {author} {\bibfnamefont {R.~W.}\ \bibnamefont
  {Boyd}},\ }\href@noop {} {\bibfield  {journal} {\bibinfo  {journal} {Optics
  letters}\ }\textbf {\bibinfo {volume} {34}},\ \bibinfo {pages} {3343}
  (\bibinfo {year} {2009})}\BibitemShut {NoStop}%
\bibitem [{\citenamefont {Wu}\ \emph {et~al.}(2011)\citenamefont {Wu},
  \citenamefont {Luo}, \citenamefont {Home}, \citenamefont {Kar},\ and\
  \citenamefont {Majumdar}}]{wulingan1}%
  \BibitemOpen
  \bibfield  {author} {\bibinfo {author} {\bibfnamefont {L.-A.}\ \bibnamefont
  {Wu}}, \bibinfo {author} {\bibfnamefont {K.-H.}\ \bibnamefont {Luo}},
  \bibinfo {author} {\bibfnamefont {D.}~\bibnamefont {Home}}, \bibinfo {author}
  {\bibfnamefont {G.}~\bibnamefont {Kar}}, \ and\ \bibinfo {author}
  {\bibfnamefont {A.~S.}\ \bibnamefont {Majumdar}},\ }in\ \href@noop {} {\emph
  {\bibinfo {booktitle} {AIP Conference Proceedings}}},\ Vol.\ \bibinfo
  {volume} {1384}\ (\bibinfo {organization} {AIP},\ \bibinfo {year} {2011})\
  pp.\ \bibinfo {pages} {223--228}\BibitemShut {NoStop}%
\bibitem [{\citenamefont {Luo}\ \emph {et~al.}(2013)\citenamefont {Luo},
  \citenamefont {Huang}, \citenamefont {Zheng},\ and\ \citenamefont
  {Wu}}]{wulingan2}%
  \BibitemOpen
  \bibfield  {author} {\bibinfo {author} {\bibfnamefont {K.~H.}\ \bibnamefont
  {Luo}}, \bibinfo {author} {\bibfnamefont {B.}~\bibnamefont {Huang}}, \bibinfo
  {author} {\bibfnamefont {W.~M.}\ \bibnamefont {Zheng}}, \ and\ \bibinfo
  {author} {\bibfnamefont {L.~A.}\ \bibnamefont {Wu}},\ }\href@noop {}
  {\bibfield  {journal} {\bibinfo  {journal} {Chinese Physics Letters}\
  }\textbf {\bibinfo {volume} {29}},\ \bibinfo {pages} {74216} (\bibinfo {year}
  {2013})}\BibitemShut {NoStop}%
\bibitem [{\citenamefont {Gao}\ \emph {et~al.}(2017)\citenamefont {Gao},
  \citenamefont {Wang}, \citenamefont {Wang}, \citenamefont {Li}, \citenamefont
  {Du}, \citenamefont {Chang},\ and\ \citenamefont {Yao}}]{Gao2017}%
  \BibitemOpen
  \bibfield  {author} {\bibinfo {author} {\bibfnamefont {C.}~\bibnamefont
  {Gao}}, \bibinfo {author} {\bibfnamefont {X.}~\bibnamefont {Wang}}, \bibinfo
  {author} {\bibfnamefont {Z.}~\bibnamefont {Wang}}, \bibinfo {author}
  {\bibfnamefont {Z.}~\bibnamefont {Li}}, \bibinfo {author} {\bibfnamefont
  {G.}~\bibnamefont {Du}}, \bibinfo {author} {\bibfnamefont {F.}~\bibnamefont
  {Chang}}, \ and\ \bibinfo {author} {\bibfnamefont {Z.}~\bibnamefont {Yao}},\
  }\href {\doibase 10.1103/PhysRevA.96.023838} {\bibfield  {journal} {\bibinfo
  {journal} {Phys. Rev. A}\ }\textbf {\bibinfo {volume} {96}},\ \bibinfo
  {pages} {023838} (\bibinfo {year} {2017})}\BibitemShut {NoStop}%
\bibitem [{\citenamefont {Shapiro}(2008)}]{cgi}%
  \BibitemOpen
  \bibfield  {author} {\bibinfo {author} {\bibfnamefont {J.~H.}\ \bibnamefont
  {Shapiro}},\ }\href@noop {} {\bibfield  {journal} {\bibinfo  {journal}
  {Physical Review A}\ }\textbf {\bibinfo {volume} {78}},\ \bibinfo {pages}
  {061802} (\bibinfo {year} {2008})}\BibitemShut {NoStop}%
\bibitem [{\citenamefont {Bromberg}, \citenamefont {Katz},\ and\ \citenamefont
  {Silberberg}(2009)}]{cgie}%
  \BibitemOpen
  \bibfield  {author} {\bibinfo {author} {\bibfnamefont {Y.}~\bibnamefont
  {Bromberg}}, \bibinfo {author} {\bibfnamefont {O.}~\bibnamefont {Katz}}, \
  and\ \bibinfo {author} {\bibfnamefont {Y.}~\bibnamefont {Silberberg}},\
  }\href@noop {} {\bibfield  {journal} {\bibinfo  {journal} {Physical Review
  A}\ }\textbf {\bibinfo {volume} {79}},\ \bibinfo {pages} {053840} (\bibinfo
  {year} {2009})}\BibitemShut {NoStop}%
\bibitem [{\citenamefont {Gong}\ and\ \citenamefont
  {Han}(2012)}]{superresolution}%
  \BibitemOpen
  \bibfield  {author} {\bibinfo {author} {\bibfnamefont {W.}~\bibnamefont
  {Gong}}\ and\ \bibinfo {author} {\bibfnamefont {S.}~\bibnamefont {Han}},\
  }\href@noop {} {\bibfield  {journal} {\bibinfo  {journal} {Physics Letters
  A}\ }\textbf {\bibinfo {volume} {376}},\ \bibinfo {pages} {1519} (\bibinfo
  {year} {2012})}\BibitemShut {NoStop}%
\bibitem [{\citenamefont {Dixon}\ \emph {et~al.}(2011)\citenamefont {Dixon},
  \citenamefont {Howland}, \citenamefont {Rodenburg}, \citenamefont {Hardy},
  \citenamefont {Simon}, \citenamefont {Sergienko}, \citenamefont {Boyd},
  \citenamefont {Chan}, \citenamefont {O'Sullivan-Hale},\ and\ \citenamefont
  {Shapiro}}]{turbulence}%
  \BibitemOpen
  \bibfield  {author} {\bibinfo {author} {\bibfnamefont {P.~B.}\ \bibnamefont
  {Dixon}}, \bibinfo {author} {\bibfnamefont {G.~A.}\ \bibnamefont {Howland}},
  \bibinfo {author} {\bibfnamefont {B.}~\bibnamefont {Rodenburg}}, \bibinfo
  {author} {\bibfnamefont {N.~D.}\ \bibnamefont {Hardy}}, \bibinfo {author}
  {\bibfnamefont {D.~S.}\ \bibnamefont {Simon}}, \bibinfo {author}
  {\bibfnamefont {A.~V.}\ \bibnamefont {Sergienko}}, \bibinfo {author}
  {\bibfnamefont {R.~W.}\ \bibnamefont {Boyd}}, \bibinfo {author}
  {\bibfnamefont {K.~W.~C.}\ \bibnamefont {Chan}}, \bibinfo {author}
  {\bibfnamefont {C.}~\bibnamefont {O'Sullivan-Hale}}, \ and\ \bibinfo {author}
  {\bibfnamefont {J.~H.}\ \bibnamefont {Shapiro}},\ }\href@noop {} {\bibfield
  {journal} {\bibinfo  {journal} {Physical Review A}\ }\textbf {\bibinfo
  {volume} {83}},\ \bibinfo {pages} {911} (\bibinfo {year} {2011})}\BibitemShut
  {NoStop}%
\bibitem [{\citenamefont {Meyers}, \citenamefont {Deacon},\ and\ \citenamefont
  {Shih}(2011)}]{turbulencei}%
  \BibitemOpen
  \bibfield  {author} {\bibinfo {author} {\bibfnamefont {R.~E.}\ \bibnamefont
  {Meyers}}, \bibinfo {author} {\bibfnamefont {K.~S.}\ \bibnamefont {Deacon}},
  \ and\ \bibinfo {author} {\bibfnamefont {Y.}~\bibnamefont {Shih}},\
  }\href@noop {} {\bibfield  {journal} {\bibinfo  {journal} {Applied Physics
  Letters}\ }\textbf {\bibinfo {volume} {98}},\ \bibinfo {pages} {111115}
  (\bibinfo {year} {2011})}\BibitemShut {NoStop}%
\bibitem [{\citenamefont {Scully}\ and\ \citenamefont
  {Zubairy}(1999)}]{quantumoptics}%
  \BibitemOpen
  \bibfield  {author} {\bibinfo {author} {\bibfnamefont {M.~O.}\ \bibnamefont
  {Scully}}\ and\ \bibinfo {author} {\bibfnamefont {M.~S.}\ \bibnamefont
  {Zubairy}},\ }\href@noop {} {\enquote {\bibinfo {title} {Quantum optics},}\ }
  (\bibinfo {year} {1999})\BibitemShut {NoStop}%
\bibitem [{\citenamefont {Katkovnik}, \citenamefont {Astola},\ and\
  \citenamefont {Egiazarian}(2008)}]{fddt}%
  \BibitemOpen
  \bibfield  {author} {\bibinfo {author} {\bibfnamefont {V.}~\bibnamefont
  {Katkovnik}}, \bibinfo {author} {\bibfnamefont {J.}~\bibnamefont {Astola}}, \
  and\ \bibinfo {author} {\bibfnamefont {K.}~\bibnamefont {Egiazarian}},\
  }\href@noop {} {\bibfield  {journal} {\bibinfo  {journal} {Applied Optics}\
  }\textbf {\bibinfo {volume} {47}},\ \bibinfo {pages} {3481} (\bibinfo {year}
  {2008})}\BibitemShut {NoStop}%
\end{thebibliography}
\end{document}